\documentclass[pre,nofootinbib,superscriptaddress]{revtex4-2}
\usepackage{parskip}
\usepackage{enumitem}

\begin{document}

\title{The Physics of Pears: A Pataphysical Approach to Geometrodynamics.}

\author{Marcello {Poletti}}
\email{epomops@gmail.com}
\affiliation{San Giovanni Bianco, Italy}

 \begin{abstract}
This work explores a pataphysical approach to the concept of space and time, inspired by Aristotelian philosophy and the insights of John Wheeler. We propose the "Physics of Pears," a theory that conceives space as a graph and time as a measure of change, within a geometrodynamical context. The goal is to stimulate reflections on the nature of physical laws and the boundaries between physics and metaphysics.
\end{abstract}

\maketitle

\section{Introduction}

In 1962, John Wheeler published \textit{Geometrodynamics} \cite{Geometrodynamics}, and on page 8, he writes:

\begin{quote}
	\textit{Is the metric continuum a magic medium which bent up in one way here represents a gravitational field, rippled in another way there describes an electromagnetic field, and twisted up locally describes a long-lived concentration of mass-energy? In other words, is physics at bottom a matter of pure geometry? Is geometry only an arena, or is it everything?}
\end{quote}

Is it possible that geometry is everything? That it is all that is required to think and describe the world? Certainly, this is not the case for Euclidean geometry and, by extension, for the space of Newtonian mechanics. This space is maximally symmetric and isotropic; all points within it are identical and interchangeable, this space is inherently empty.

However, Riemannian geometry has opened the door to wonderfully intricate spaces, maximally asymmetric and anisotropic, in which every portion, every neighborhood of every point, can carry a variety of information — these varieties could be what we call "mass", "field", or "energy."

This program has, in a sense, been successfully completed by general relativity. In this domain, geometry is more properly pseudo-geometry, but the flavor of Geometrodynamics is unmistakable. The field equations \cite{Einstein1} can indeed be read almost literally following Wheeler's quote: the folds of the pseudo-metric $g_{\mu\nu}$ are equivalent to (correspond to, describe) a certain distribution of mass and energy $T_{\mu\nu}$ via the Einstein tensor $G_{\mu\nu}$.

Shortly after Einstein's publication, Theodor Kaluza \cite{Kaluza} extended the mechanism to electromagnetism, and all the forces known at the time could be interpreted geometrically.

There are many complications; Einstein's geometry is not strictly geometry due to the signature of his tensors, and Kaluza's hypothesis requires an additional mysterious spatial dimension.

Here, a toy theory called "The Physics of Pears" will be described. The playful name underscores the lighthearted (pataphysical!) nature of the operation, and the pears specifically refer to the fact that the theory is strictly finitistic, made up of integers and counting, much like counting pears or crates of pears.

This toy theory offers little in terms of formal rigor; however, as we shall see, it contains three ideas that may have value, at least philosophically.

The first of these ideas [Chapter 2, \textit{The Space of Pears}] is that the concept of "number of dimensions" may be less significant than it seems. From a geometrodynamic perspective, it is possible to work with an arbitrary number of dimensions without worrying too much about adhering to real physical space and without attempting to conceal the extra dimensions through compactification \cite{Klein}.

The point is that "real" physical space is "obviously three-dimensional" due to a series of simple experiments anyone can perform, particularly, there is no way to weld four iron rods together in such a way that they are all mutually perpendicular. Three is possible, but four is not. However, fully accepting Wheeler's provocation, the iron rods, strictly speaking, do not exist but are manifestations emerging from the underlying geometry.

The Physics of Pears proposes a much simpler and more elementary geometric model than the usual continuous varieties and shows how, from these geometries, a continuous three-dimensional representation always emerges. In other words, this theory suggests a line of thought that allows us to think of matter as the "three-dimensional interpretation" of an underlying dimensionless geometry.

The second idea concerns time [Chapter 3, \textit{The Time of Pears}]. On this debated subject, the Physics of Pears adopts a concrete and operational perspective, reviving Aristotle's notion of time, time as the measure of change. This notion is simple, deeply satisfying on a philosophical level, and leads very naturally to conclusions closely aligned with the time of Einstein's relativity. The proposal, seemingly innocuous, upends much of physics, which is oriented towards studying quantities like ${d\mathcal{F}}/{dt}$, i.e., studying how things change "while" time changes. In contrast, the line of interpretation here based on Aristotle's thinking posits that, at most, things change and that "time" is the measure of this change. If $\mathcal{F}$ is "everything that changes", at most we have $dt=k \, d\mathcal{F}$, where $k$ is constant. This seemingly absurd idea leads to surprising consequences.

A brief chapter on pear cosmology [Chapter 4, \textit{Cosmology of the Pears}] will venture hypotheses on the history of a pear universe. It is likely the most pataphysical, forced, and pseudo-scientific chapter, and readers who do not enjoy the wildest speculation are advised to skip it.

The following chapter [Chapter 5, \textit{Metaphysics of the Pears}] addresses the relationship between the Physics of Pears and quantum mechanics, and, more generally, the relationship between metaphysics and physics. There, the third idea will be discussed: the hypothesis that quantum mechanics may be correct and non-completable, yet simultaneously absent from metaphysics and, in some sense, incomplete. The idea is that QM is primarily about physics, predicting measurement outcomes, and says little about the ontology of the world's things.

\section{The Space of Pears}

In the 1600s, thanks particularly to the work of François Viète \cite{Viete} and René Descartes \cite{Descartes}, analytic geometry was born and became established, offering a revolutionary approach to solving geometric problems. Rather than referring to intrinsic geometric properties, such as distances and angles, typical of Euclidean geometry, analytic geometry introduces the use of coordinates, or ordered pairs of numbers that allow the description of a point's position in the plane. Geometric properties can then be reconstructed a posteriori as functions of the coordinates themselves.

For example, the distance \( d(P_1, P_2) \) between two points \( P_1 \) and \( P_2 \) in a Cartesian plane takes the form:

\begin{equation}
d(P_1, P_2) = \sqrt{(x_1 - x_2)^2 + (y_1 - y_2)^2} \label{eq:dist}
\end{equation}

Where the pairs \( (x_1, y_1) \) and \( (x_2, y_2) \) are the coordinates associated with \( P_1 \) and \( P_2 \), respectively.

The concept of distance is critical in geometry (and therefore in analytic geometry), and it is directly connected to the Pythagorean theorem (and hence to the \ref{eq:dist} and to the fifth discussed parallel postulate).

Given two points \( P_1 \) and \( P_2 \), the distance \( d(P_1, P_2) = d(P_2, P_1) \) is an arbitrary real value, non-negative, and equal to 0 if and only if \( P_1 \equiv P_2 \). By adding a third point \( P_3 \), the concept of distance is structurally defined by the triangle inequality:

\begin{equation}
d(P_1, P_2) + d(P_2, P_3) \geq d(P_1, P_3) \label{eq:tri}
\end{equation}

Equation \ref{eq:tri} captures and defines the intuitive idea that "the distance between A and B is the length of the shortest path between A and B," and it serves as a constraint on the possible relative values of distance. 

Consider the following hypothetical start to a geometry problem: "Given a triangle with sides \(a=1, b=2, c=4\)...", this problem is clearly poorly formulated, or contains a trap, as it assumes an impossible triangle as the starting condition. That being said, a Euclidean triangle has no other constraints; that is, any set of lengths that satisfies equation \ref{eq:tri} represents a valid triangle.\footnote{It is worth noting how analytic geometry brilliantly overcomes this minor difficulty. The start "Given a triangle with coordinates a,b,c..." never contains any ambiguity, as Equation 1 is formulated in such a way that it always guarantees the satisfaction of Equation \ref{eq:tri}.}

Now, considering a fourth point \( P_4 \), the problem becomes more complex, and the concept of "number of dimensions" takes on a role in defining constraints on distances. 

Let us consider four points \( P_1, P_2, P_3, P_4 \) such that:

\begin{equation}
d(P_1, P_2) = d(P_1, P_3) = d(P_2, P_3) = d(P_2, P_4) = d(P_3, P_4) = 1 \label{eq:xxx}
\end{equation}

These distances satisfy Equation \ref{eq:tri} and are therefore valid. The question now is what values the single missing distance \( d(P_1, P_4) \) can take. In Euclidean planar geometry, there are only two possibilities: either \( P_1 \equiv P_4 \), and therefore \( d(P_1, P_4) = 0 \), or \( P_1 \not\equiv P_4 \), and \( d(P_1, P_4) = \sqrt{3} \). In analytic geometry, this corresponds to the fact that any set of coordinates that satisfies Equation \ref{eq:xxx} results in a value of \( d(P_1, P_4) \in (0, \sqrt{3}) \). 

This constraint, unlike Equation 2, is specific to planar geometry; in solid geometry, it loosens, allowing \( P_1 P_4 \) to take any value within the interval \([0, \sqrt{3}]\).

The number of dimensions, in this sense, operates as a constraint that, like Equation 2, limits the allowable distances.

On an entirely different front, non-Euclidean geometries, in terms of distances, operate in the same way. They allow distances not permitted in "flat" geometry, violating the Pythagorean theorem.

Non-Euclidean geometries indeed allow alternative forms of Equation 1, but always within the bounds of Equation 2. This flexibility corresponds to possible non-Euclidean distances. For instance, in spherical geometry, it is allowed that \( \overline{P_1 P_4} = 2 \).

The addition of dimensions and non-Euclidean geometries work together to make the concept of distance more flexible. A Riemannian manifold of appropriate dimension allows access to distances between points that are solely constrained by the triangle inequalities.

These observations lead to a strong suggestion: is it possible to think of physical space in terms of distance relations governed by Equation \ref{eq:tri}, without any further constraints? This suggestion directly leads to Mach's attempt \cite{Mach} to think of space in terms of distances between bodies, i.e., physical space as the space of distance relations.

Mathematically, analytic geometry was formalized in the concept of a normed vector space, which captures all the aspects discussed here. The suggested geometric space corresponds to the weaker mathematical concept of a metric space. The suggestion can therefore be formulated as follows: is it possible that physical space should be thought of as a metric space rather than as a conventional normed vector space?

The price to pay is high, as in all known physics, physical space is modeled through vector spaces (or tensor extensions in the case of general relativity), but the wager is appealing: a space that is neither Euclidean nor non-Euclidean, that does not immediately define "dimensions," that does not contain a clear definition of curvature, but simply defines distances (any) between points.

But what kind of geometry can be done with such an unstructured concept as that of a metric space? Consider that a metric space does not even provide a natural definition of angle \cite{Mondino}, surface, or volume.

In the game of the Physics of Pears, a truly elementary form of metric space will be used: undirected graphs with distances defined by the shortest path. Graphs are more properly finite proximity spaces, sets equipped with the symmetric notion "A is proximal to B" and on which distance can be simply defined as the minimal length of the possible proximity chains. Unlike Mach, therefore, the Physics of Pears does not think of space as an emerging relational structure, but rather as an object, with an inherent nature (as in Newton, Einstein, and Geometrodynamics). The space of pears is a finite set of points that can, in pairs, either be or not be, proximal to each other.

A space of pears has a characteristic entirely absent in conventional physical/mathematical space and even in usual metric spaces: it can be connected or disconnected, meaning it is possible that no chain of proximity exists between A and B, and therefore the distance d(A,B) is undefined. Unless otherwise specified, the following discussion will refer only to connected graphs. 

For a (connected) graph, the notion of distance is defined for every pair of points, and the maximum of these distances is called the diameter.

A well-known class of graphs, planar graphs, has an interesting connection with the considerations above. A graph is said to be planar if it can be represented on a plane without intersection of edges. Once such a representation is produced, the plane can be deformed so that the plane's metric coincides with the metric of the graph. 

In other words:

\begin{itemize}[label=\textendash]
	\item A planar graph can be approximated by an appropriate curved two-dimensional manifold.
\end{itemize}

Various types of lattice structures are used in different fields to approximate continuous spaces, here the opposite is being done, as planar graphs are being approximated by continuous two-dimensional varieties. The concepts of curvature and two-dimensionality, therefore, can be thought of as emerging. A planar graph does not expose any natural notion of surface or angle, yet such notions are present in continuous spaces that approximate the graph's metric. \footnote{It should also be noted that the connection between planar graph and 2D manifold is somewhat fuzzy and depends on how the graph is represented on the plane.}

One may naturally wonder how this result can be generalized. Indeed, similarly, any graph can be represented in a three-dimensional Euclidean space without intersection of edges. Once such a representation is produced, the Euclidean space can be deformed so that the manifold's metric coincides with the metric of the graph. \cite{SpaceAsRelation}

In other words:

\begin{itemize}[label=\textendash]
	\item A generic graph can be approximated by an appropriate three-dimensional curved manifold.
\end{itemize}

This is quite surprising. On one hand, it suggests that the evolution of geometry from Euclidean to non-Euclidean, and up to the study of Riemannian manifolds, can be reinterpreted as a progressive attempt to extend the metric possibilities by removing constraints other than triangle inequalities. On the other hand, it shows how three dimensions represent the most economical Euclidean curved space for making this type of continuous representation.

The interesting fact about the space of pears is not that "it explains why spatial dimensions are exactly three", but the deeper fact that it exemplifies how dimensions and continuity [both Euclidean and non-Euclidean geometries] can emerge as accidents from a structure that lacks these concepts. If physical space were a space of pears, it would be entirely natural for an appropriately sentient agent to categorize it as a roughly curved 3D space.

Without hesitation, the geometrodynamic hypothesis:

\begin{itemize}[label=\textendash]
	\item A space of pears can be approximately categorized as a three-dimensional space containing matter and energy.
\end{itemize}

It's a bit crazy, but it's the game of pears.

\section{The Time of Pears}

The debate about time is vast, encompassing almost every possible position, from "Time does not exist" \cite{Rovelli} to "Only time exists" \cite{Smolin}. Personally, I find the Aristotelian concept of time \cite{Aristotele} as the measure of change philosophically very satisfying. The idea is that things do not change in time, but that things change, and the measure of this change is called time.

This conception is in direct contrast to Newton's view, as it does not, by definition, admit that "a body persists in its state of rest," meaning that nothing changes and time continues to flow.

It is interesting (and amusing) to apply this idea to the elementary mechanics of a material point. In this abstraction, the only variable subject to change is position, which leads directly to the definition:

\begin{equation}
\Delta T = \frac{\Delta S}{c} \label{eq:time}
\end{equation}

where \( c \) is a constant with the dimensions of a velocity that only serves to convert units. In differential form:

\begin{equation}
dx^2 + dy^2 + dz^2 - c^2 dt^2 = 0 \label{eq:timedif}
\end{equation}

This is the metric for light-like paths in a Minkowski space.

To model bodies moving at speeds other than c, one must assume that the body has some kind of "internal motion." Consider, for example, a material point moving in a circular path around a fixed center in a Euclidean plane. This circular object can be regarded overall as an extended body capable of persisting in a state of rest.

Notice how this image is not far from the usual one of a Newtonian body at rest, which we all imagine as "at rest" even though it is, of course, composed of atoms and molecules that vibrate, oscillate, and are generally in continuous motion.

If such an object is set in motion, part of the displacement of the material point of which it is composed will be dedicated to translation rather than circular motion, so that:

\begin{itemize}[label=\textendash]
	\item The body cannot, by definition, reach or exceed velocity c
	\item Internal motion is slowed in the presence of external motion.
\end{itemize}

A few steps \cite{TimeAsChange} are sufficient to show that the motion of the body, as defined here, perfectly respects the Lorentz transformations. Aristotelian time is spontaneously relativistic. Additionally, the following powerful suggestions emerge:

\begin{itemize}[label=\textendash]
	\item Mass is a form of motion, or internal energy.
	\item Only massless objects can travel at speed c, and such objects can only travel at speed c by definition, as shown in Equation \ref{eq:time}.
\end{itemize}

It is certainly vague, but undeniably troubling: the twin left on Earth is at rest and "uses" all its changes, i.e., its time, for internal motion, to live and age. When it reunites with the space-traveling twin, it finds them young, with a youthful memory and a youthful clock that has ticked much less than their own. This is due to the fact that the younger twin "used" much of their changes for travel and only the remainder for aging and ticking.\footnote{Note that the relationship is in any case relative. That is, the twin at rest is a state of rest relative to the space-traveling twin, but both can be in motion relative to a third party.}

Beyond the suggestions, Equations \ref{eq:time} and \ref{eq:timedif}, combined with the Aristotelian perspective, lead to a more radical vision of the usual Minkowski space-time. Space and time may not simply be interconnected, but one may define the other. In particular, relativistic time can be read as the measure of spatial changes.

Now, we need to make some observations regarding the application of Aristotelian time to classical elementary mechanics in the case of more than one body. Let us consider two bodies at rest, represented by two material points in circular motion. It is not at all clear how these two bodies are or are not synchronized. Suppose that, by chance, a change occurs in body 1 or body 2, with a symmetric probability distribution such that the two bodies are, on average, synchronized. There are three different points of view, corresponding to three different "times":

\begin{itemize}[label=\textendash]
	\item Considering the pair of bodies as one entity, there is a tick of time at each change, whether it occurs in body 1 or body 2.
	\item Considering body 1, time only passes when it undergoes a change.
	\item Similarly, from the perspective of body 2, time only passes when it undergoes a change.
\end{itemize}

Thus, time, from a certain perspective, is a strictly local fact. A change in A is a tick of time for A, or for the pair AB, but it is not a tick of time for B.

This kind of time has a thermodynamic flavor. This time can be summed globally [as if obtaining a total energy, which increases with the number of bodies] or it can be redefined as an average over the bodies themselves, yielding something akin to temperature, which does not increase or decrease by adding or removing a molecule of average energy. Operating in terms of averages introduces imprecision, but it brings us closer to the usual idea of time, where for body A, time has passed by x, the same for body B, and similarly for the pair AB, the time passed is x, not 2x.

In the Physics of Pears, this approach will be adopted with a certain lightness, in an elementary form:

\begin{itemize}[label=\textendash]
	\item Given two graphs, isomorphic except for the addition or removal of an edge, they will be said to be one temporal unit apart.
\end{itemize}

We will also make a second arbitrary assumption:

\begin{itemize}[label=\textendash]
	\item The removal of an edge is one unit step into the future.
	\item The addition of an edge is one unit step into the past.
\end{itemize}

A space of pears, then, changes in minimal steps consisting of the removal of an edge. Despite the fact that the Physics of Pears postulates some sort of absolute space (the graph) and an absolute time (the inexorable loss of edges), there seems to be a strong possibility that the Physics of Pears is Lorentz invariant and that this is generally true for a theory that properly embraces the Aristotelian view of time.

In the Physics of Pears, there is no problem of multi-bodies as emerged above, since there are no bodies involved, and the entire universe is treated as a whole. However, it remains true that temporal evolution concerns the universe as a whole and those subgraphs that contain the removed edges, but not those subgraphs that do not.

\section{Cosmology of the Pears}

A pear universe is, therefore, a graph that evolves by the loss of edges, nothing more. Due to the removal of edges, in general, the diameter of the graph increases as the possible paths between two points decrease. The graph expands, and the rate of expansion itself increases. The graph expands with accelerated motion.

However, this process is hindered by the phenomenon of disconnection. Occasionally, the loss of an edge may cause a point or an entire subgraph to become disconnected; we will refer to this phenomenon as evaporation. Following the evolution of the largest connected subgraph, it expands at increasing speeds until the rate of evaporation becomes dominant. At this point, the diameter decreases rapidly until the graph evaporates completely. The ultimate future of any pear universe is inescapably determined: a set of disconnected points.

By reversing the reasoning, it is possible to estimate the origin of a pear universe U, though with many limitations. The progressive addition of edges leads to the conclusion that the origin of U consists of a complete graph, with diameter 1, perfectly symmetric and uniform, and therefore in an initial state of minimal size, high density, and low entropy. However, the phenomenon of evaporation complicates the picture, preventing a precise placement of this origin in time. Evaporation tears information from the graph, entropy increases (until the evaporation phase dominates), and this makes the past largely unpredictable. A pear universe evolves, carrying with it traces of its own evolution, but the phenomenon of evaporation increasingly erases these very traces.

Thus, a pear universe begins with a Big Bang, a state in which the graph has maximum density, minimum diameter, and minimum entropy. It expands with an increasing rate of expansion, but at the same time, the evaporation rate increases, opposing the expansion. The two processes eventually reverse, with the evaporation rate surpassing the growth rate, and the universe dissolves into a cloud of disconnected points.

\section{Metaphysics of Pears}

The Physics of Pears is a metaphysical toy theory. It speculates on how the world is, not necessarily on how it functions. Naturally, physics and metaphysics intertwine. A good metaphysics guides the physicist in the search for good physical recipes, and conversely, a physical theory leads to a worldview—Newton's absolute space and time, the impalpable but immanent fields of Faraday and Maxwell, and Einstein's "mollusk" spacetime.

Then, there is quantum mechanics, which brings our desire to visualize the world to its knees, providing us with bizarre representations, from the many worlds to the many minds.

However, it must be noted that physics and metaphysics are not, and cannot be, particularly adjacent. Metaphysics offers a perspective from the point of view of the deus ex machina, observing reality without interaction, for what it is. Physics, on the other hand, makes predictions about what one part of the world, the experimenter, measures in relation to another part, the object of experiment. Historically, physics has operated in a metaphysical way, describing the systems it is interested in from the standpoint of a deus ex machina, who is not part of the system itself. Physics has been historically developed under the assumption that the role of the physicist is negligible, but this is not the case.

Consider the following situation: a system A (for instance, a person, Alice) is separated from a system B (Bob). Which of the following two propositions best describes physical reality?

\begin{itemize}[label=\textendash]
	\item System A contains some information about system B [Alice knows Bob], but this does not mean that B is fully projected into A, that all possible information about B, past and future, is available in A, or that B is a holographic projection of A.
	\item System B is redundant because all information about the entire history of the universe is available in every arbitrarily small spacetime neighborhood, and therefore in A.
\end{itemize}

Position 1 seems the most obvious, prudent, and natural. Position 2 seems an absurd a priori assumption: a grain of sand will carry many signs of its history, much information about what is around it, but not necessarily the entire universal knowledge.

Yet, incredibly, classical physics, which leaves us more at ease than quantum mechanics on the ontological level, adopts position 2. The history of the motion of even the tiniest particle for a tiny period of time depends on the entire amount of mass and fields in the universe, according to formulas so analytically precise that the entire universe, in its entirety, can be thought of as the analytic extension of that particle. This is a radical version of Laplace's demon \cite{Strangeness}, leading inevitably to the most radical solipsism: the universe is the extension of my Cartesian ego.

Realism, in its most elementary form, comes from the negation of solipsism: "There is something beyond my Cartesian ego," and this existence should not be a fideistic choice but the acknowledgment that something exists because it is not the complete extension of my Cartesian ego, that is, something exists because it is not fully in me.

Alice knows Bob, she has many traces of Bob within her, she knows his appearance, remembers his accent and character, but there are aspects of Bob that Alice cannot know, deeply unknown to her, with no trace in Alice.

This is such a simple and obvious position that it leads us to shift all ontological doubts regarding quantum mechanics to classical physics: how can physics consider that all the information of the universe, past, present, and future, is enclosed in the trajectory, and its derivatives, of a material point for a second? It is nonsense or, at the very least, it would be truly surprising if things were this way.

But if Bob has a property P that is entirely and profoundly undecidable for Alice, what kind of physics can Alice produce? Alice will only be able to make probabilistic guesses about the values of P, and this opens up a problem: what kind of probabilities?

Classical probability is defined in classical contexts and perfectly handles classical cases (such as dice, coins, votes...), but it is not necessarily the most suitable mathematical framework to handle the unknowability we are discussing. And indeed, one can argue \cite{Strangeness2} that this type of unknowability should be dealt with by different tools, specifically the first postulates of quantum mechanics. Quantum mechanics violates Bell's inequalities \cite{Bell} not because the objects have a strange ontology, but because that kind of epistemic lack of knowledge is different, deeper, than the usual dice toss and requires a different theory, which violates classical probability theory to the extent identified by Bell's inequalities.

Thus, Alice will treat Bob's property P using quantum mechanics, but without quantum mechanics saying anything about the ontology of P. And conversely, Bob will do the same with some properties P' of Alice. Bob is a "quantum object" for Alice as much as Alice is for Bob. The experimenter is in a state of superposition from the perspective of the cat \cite{Schrodinger} as much as the cat is from the experimenter's point of view, and both are so for Wigner \cite{Wigner}.

Physics is tasked with making predictions about measurements, not describing reality. The implicit classical hypothesis was that these two things coincided, but this was plainly a forced assumption. Making predictions means constructing the appropriate mathematical tools to handle a scarcity of knowledge, and these tools are quantum mechanics.

Consequently, if we imagine a metaphysics, it is unnecessary to "impose quantum structures" on it. If this metaphysics is not "analytic" like classical physics, then the physics that emerges from it, the physics that produces a physicist within the universe of pears, will inevitably be mediated by quantum probabilities.

\section{Conclusions}

The Physics of Pears is a haphazard game. It will likely irritate those who draw a sharp line between pseudo-science and real-science. But it is a game that sometimes needs to be played, and perhaps, amid the folds of delirium, there is something worth saying, something that could stimulate further reflection, other ideas, even negatively. 

And deep down, in our most secret thoughts, I hope it leaves at least some concrete doubt: what if we are all really just pulp in a great pear juice?

\section{Acknowledgments}

A thank you goes to Andrea Martinelli, who, thirty years ago, on the Milano-Bergamo train, back and forth from university, actively wasted his time with me on these ideas. 

Any clearly wrong concept is to be attributed to him, any good or decent thing to me.

A special thanks goes to Carlo Rovelli for being one of the greatest experts in the world on the Physics of Pears, having endured and tolerated with a patience bordering on sainthood a number of hypotheses, partial writings, and deliriums, of which this work is a sober summary.

A special thanks also to ChatGPT, which acted as a mirror during the writing process and contributed to the touch of madness in this work. Additionally, it professionally handled the English writing, discussing and suggesting the best linguistic variants in every aspect.
Finally, ChatGPT wrote, entirely with its own hands, the abstract.

\nocite{*}

\bibliography{Eng}% Produces the bibliography via BibTeX.

%apsrev4-2.bst 2019-01-14 (MD) hand-edited version of apsrev4-1.bst
%Control: key (0)
%Control: author (8) initials jnrlst
%Control: editor formatted (1) identically to author
%Control: production of article title (0) allowed
%Control: page (0) single
%Control: year (1) truncated
%Control: production of eprint (0) enabled
\begin{thebibliography}{18}%
\makeatletter
\providecommand \@ifxundefined [1]{%
 \@ifx{#1\undefined}
}%
\providecommand \@ifnum [1]{%
 \ifnum #1\expandafter \@firstoftwo
 \else \expandafter \@secondoftwo
 \fi
}%
\providecommand \@ifx [1]{%
 \ifx #1\expandafter \@firstoftwo
 \else \expandafter \@secondoftwo
 \fi
}%
\providecommand \natexlab [1]{#1}%
\providecommand \enquote  [1]{``#1''}%
\providecommand \bibnamefont  [1]{#1}%
\providecommand \bibfnamefont [1]{#1}%
\providecommand \citenamefont [1]{#1}%
\providecommand \href@noop [0]{\@secondoftwo}%
\providecommand \href [0]{\begingroup \@sanitize@url \@href}%
\providecommand \@href[1]{\@@startlink{#1}\@@href}%
\providecommand \@@href[1]{\endgroup#1\@@endlink}%
\providecommand \@sanitize@url [0]{\catcode `\\12\catcode `\$12\catcode
  `\&12\catcode `\#12\catcode `\^12\catcode `\_12\catcode `\%12\relax}%
\providecommand \@@startlink[1]{}%
\providecommand \@@endlink[0]{}%
\providecommand \url  [0]{\begingroup\@sanitize@url \@url }%
\providecommand \@url [1]{\endgroup\@href {#1}{\urlprefix }}%
\providecommand \urlprefix  [0]{URL }%
\providecommand \Eprint [0]{\href }%
\providecommand \doibase [0]{https://doi.org/}%
\providecommand \selectlanguage [0]{\@gobble}%
\providecommand \bibinfo  [0]{\@secondoftwo}%
\providecommand \bibfield  [0]{\@secondoftwo}%
\providecommand \translation [1]{[#1]}%
\providecommand \BibitemOpen [0]{}%
\providecommand \bibitemStop [0]{}%
\providecommand \bibitemNoStop [0]{.\EOS\space}%
\providecommand \EOS [0]{\spacefactor3000\relax}%
\providecommand \BibitemShut  [1]{\csname bibitem#1\endcsname}%
\let\auto@bib@innerbib\@empty
%</preamble>
\bibitem [{\citenamefont {Wheeler}(1962)}]{Geometrodynamics}%
  \BibitemOpen
  \bibfield  {author} {\bibinfo {author} {\bibfnamefont {J.~A.}\ \bibnamefont
  {Wheeler}},\ }\href@noop {} {\emph {\bibinfo {title} {Geometrodynamics}}}\
  (\bibinfo  {publisher} {Academic Press},\ \bibinfo {address} {New York},\
  \bibinfo {year} {1962})\BibitemShut {NoStop}%
\bibitem [{\citenamefont {Einstein}(1915)}]{Einstein1}%
  \BibitemOpen
  \bibfield  {author} {\bibinfo {author} {\bibfnamefont {A.}~\bibnamefont
  {Einstein}},\ }\bibfield  {title} {\bibinfo {title} {Die feldgleichungen der
  gravitation},\ }\href@noop {} {\bibfield  {journal} {\bibinfo  {journal}
  {Sitzungsberichte der Königlich Preußischen Akademie der Wissenschaften
  (Berlin)}\ ,\ \bibinfo {pages} {844}} (\bibinfo {year} {1915})}\BibitemShut
  {NoStop}%
\bibitem [{\citenamefont {Kaluza}(1921)}]{Kaluza}%
  \BibitemOpen
  \bibfield  {author} {\bibinfo {author} {\bibfnamefont {T.}~\bibnamefont
  {Kaluza}},\ }\bibfield  {title} {\bibinfo {title} {Zum unitätsproblem in der
  physik},\ }\href@noop {} {\bibfield  {journal} {\bibinfo  {journal}
  {Sitzungsberichte der Königlich Preußischen Akademie der Wissenschaften
  (Berlin)}\ ,\ \bibinfo {pages} {966}} (\bibinfo {year} {1921})}\BibitemShut
  {NoStop}%
\bibitem [{\citenamefont {Klein}(1926)}]{Klein}%
  \BibitemOpen
  \bibfield  {author} {\bibinfo {author} {\bibfnamefont {O.}~\bibnamefont
  {Klein}},\ }\bibfield  {title} {\bibinfo {title} {Quantum theory and
  five-dimensional theory of relativity},\ }\href@noop {} {\bibfield  {journal}
  {\bibinfo  {journal} {Zeitschrift für Physik}\ }\textbf {\bibinfo {volume}
  {37}},\ \bibinfo {pages} {895} (\bibinfo {year} {1926})}\BibitemShut
  {NoStop}%
\bibitem [{\citenamefont {Viète}(1591)}]{Viete}%
  \BibitemOpen
  \bibfield  {author} {\bibinfo {author} {\bibfnamefont {F.}~\bibnamefont
  {Viète}},\ }\href@noop {} {\emph {\bibinfo {title} {In Artem Analyticam
  Isagoge}}}\ (\bibinfo  {publisher} {Typographia Plantiniana},\ \bibinfo
  {year} {1591})\ \bibinfo {note} {geometria, quae prius solebat verborum
  interpretatione, nunc verorum numerorum calculatione tractanda
  est.}\BibitemShut {Stop}%
\bibitem [{\citenamefont {Descartes}(1637)}]{Descartes}%
  \BibitemOpen
  \bibfield  {author} {\bibinfo {author} {\bibfnamefont {R.}~\bibnamefont
  {Descartes}},\ }\href@noop {} {\emph {\bibinfo {title} {La Géométrie}}}\
  (\bibinfo  {publisher} {Imprimerie de la Compagnie des Bibliophiles},\
  \bibinfo {year} {1637})\ \bibinfo {note} {si on savait comment tracer un
  point par ses deux dimensions, on pourrait aussi bien tracer la figure de la
  même manière.}\BibitemShut {Stop}%
\bibitem [{\citenamefont {Mach}(1907)}]{Mach}%
  \BibitemOpen
  \bibfield  {author} {\bibinfo {author} {\bibfnamefont {E.}~\bibnamefont
  {Mach}},\ }\href@noop {} {\emph {\bibinfo {title} {The Science of Mechanics:
  A Critical and Historical Account of Its Development}}}\ (\bibinfo
  {publisher} {Open Court Publishing},\ \bibinfo {address} {Chicago},\ \bibinfo
  {year} {1907})\ \bibinfo {note} {translated by Thomas J.
  McCormack}\BibitemShut {NoStop}%
\bibitem [{\citenamefont {Han}\ and\ \citenamefont {Mondino}(2017)}]{Mondino}%
  \BibitemOpen
  \bibfield  {author} {\bibinfo {author} {\bibfnamefont {B.-X.}\ \bibnamefont
  {Han}}\ and\ \bibinfo {author} {\bibfnamefont {A.}~\bibnamefont {Mondino}},\
  }\bibfield  {title} {\bibinfo {title} {Angles between curves in metric
  measure spaces},\ }\href@noop {} {\bibfield  {journal} {\bibinfo  {journal}
  {Analysis and Geometry in Metric Spaces}\ }\textbf {\bibinfo {volume} {5}},\
  \bibinfo {pages} {47} (\bibinfo {year} {2017})}\BibitemShut {NoStop}%
\bibitem [{\citenamefont {Poletti}(2022{\natexlab{a}})}]{SpaceAsRelation}%
  \BibitemOpen
  \bibfield  {author} {\bibinfo {author} {\bibfnamefont {M.}~\bibnamefont
  {Poletti}},\ }\bibfield  {title} {\bibinfo {title} {Space as relation},\
  }\href {https://arxiv.org/abs/2202.02985} {\bibfield  {journal} {\bibinfo
  {journal} {arXiv preprint arXiv:2202.02985}\ } (\bibinfo {year}
  {2022}{\natexlab{a}})}\BibitemShut {NoStop}%
\bibitem [{\citenamefont {Rovelli}(2009)}]{Rovelli}%
  \BibitemOpen
  \bibfield  {author} {\bibinfo {author} {\bibfnamefont {C.}~\bibnamefont
  {Rovelli}},\ }\bibfield  {title} {\bibinfo {title} {Forget time},\
  }\href@noop {} {\bibfield  {journal} {\bibinfo  {journal} {Arxiv:
  gr-qc/0904.0675}\ } (\bibinfo {year} {2009})}\BibitemShut {NoStop}%
\bibitem [{\citenamefont {Smolin}(2013)}]{Smolin}%
  \BibitemOpen
  \bibfield  {author} {\bibinfo {author} {\bibfnamefont {L.}~\bibnamefont
  {Smolin}},\ }\href@noop {} {\emph {\bibinfo {title} {Time Reborn: From the
  Crisis in Physics to the Future of the Universe}}}\ (\bibinfo  {publisher}
  {Houghton Mifflin Harcourt},\ \bibinfo {year} {2013})\BibitemShut {NoStop}%
\bibitem [{\citenamefont {Aristotele}(1981)}]{Aristotele}%
  \BibitemOpen
  \bibfield  {author} {\bibinfo {author} {\bibnamefont {Aristotele}},\
  }\href@noop {} {\emph {\bibinfo {title} {Fisica IV}}}\ (\bibinfo  {publisher}
  {Tradotto da: Carlo Diano},\ \bibinfo {year} {1981})\ \bibinfo {note} {testo
  greco a cura di Giorgio Reale, traduzione italiana, Milano:
  Bompiani}\BibitemShut {NoStop}%
\bibitem [{\citenamefont {Poletti}(2022{\natexlab{b}})}]{TimeAsChange}%
  \BibitemOpen
  \bibfield  {author} {\bibinfo {author} {\bibfnamefont {M.}~\bibnamefont
  {Poletti}},\ }\bibfield  {title} {\bibinfo {title} {Time as change},\ }\href
  {https://arxiv.org/abs/2201.01944} {\bibfield  {journal} {\bibinfo  {journal}
  {arXiv preprint arXiv:2201.01944}\ } (\bibinfo {year}
  {2022}{\natexlab{b}})}\BibitemShut {NoStop}%
\bibitem [{\citenamefont {Poletti}(2022{\natexlab{c}})}]{Strangeness}%
  \BibitemOpen
  \bibfield  {author} {\bibinfo {author} {\bibfnamefont {M.}~\bibnamefont
  {Poletti}},\ }\bibfield  {title} {\bibinfo {title} {On the strangeness of
  quantum mechanics},\ }\href {https://doi.org/10.1007/s10701-022-00582-w}
  {\bibfield  {journal} {\bibinfo  {journal} {Foundations of Physics}\ }\textbf
  {\bibinfo {volume} {52}},\ \bibinfo {pages} {1} (\bibinfo {year}
  {2022}{\natexlab{c}})}\BibitemShut {NoStop}%
\bibitem [{\citenamefont {Poletti}(2023)}]{Strangeness2}%
  \BibitemOpen
  \bibfield  {author} {\bibinfo {author} {\bibfnamefont {M.}~\bibnamefont
  {Poletti}},\ }\bibfield  {title} {\bibinfo {title} {On the strangeness of
  quantum probabilities},\ }\href {https://doi.org/10.1007/s40509-023-00299-z}
  {\bibfield  {journal} {\bibinfo  {journal} {Quantum Studies: Mathematics and
  Foundations}\ }\textbf {\bibinfo {volume} {10}},\ \bibinfo {pages} {343}
  (\bibinfo {year} {2023})}\BibitemShut {NoStop}%
\bibitem [{\citenamefont {Bell}(1964)}]{Bell}%
  \BibitemOpen
  \bibfield  {author} {\bibinfo {author} {\bibfnamefont {J.~S.}\ \bibnamefont
  {Bell}},\ }\bibfield  {title} {\bibinfo {title} {On the einstein podolsky
  rosen paradox},\ }\href {https://doi.org/10.1103/PhysicsPhysiqueFizika.1.195}
  {\bibfield  {journal} {\bibinfo  {journal} {Physics Physique Fizika}\
  }\textbf {\bibinfo {volume} {1}},\ \bibinfo {pages} {195} (\bibinfo {year}
  {1964})}\BibitemShut {NoStop}%
\bibitem [{\citenamefont {Schrödinger}(1935)}]{Schrodinger}%
  \BibitemOpen
  \bibfield  {author} {\bibinfo {author} {\bibfnamefont {E.}~\bibnamefont
  {Schrödinger}},\ }\bibfield  {title} {\bibinfo {title} {Die gegenwärtige
  situation in der quantenmechanik},\ }\href
  {https://doi.org/10.1007/BF01491891} {\bibfield  {journal} {\bibinfo
  {journal} {Naturwissenschaften}\ }\textbf {\bibinfo {volume} {23}},\ \bibinfo
  {pages} {807} (\bibinfo {year} {1935})}\BibitemShut {NoStop}%
\bibitem [{\citenamefont {Wigner}(1961)}]{Wigner}%
  \BibitemOpen
  \bibfield  {author} {\bibinfo {author} {\bibfnamefont {E.~P.}\ \bibnamefont
  {Wigner}},\ }\bibfield  {title} {\bibinfo {title} {Remarks on the mind-body
  question},\ }\href {https://doi.org/10.1080/00318161.1961.10545780}
  {\bibfield  {journal} {\bibinfo  {journal} {Philosophical Magazine}\ }\textbf
  {\bibinfo {volume} {46}},\ \bibinfo {pages} {1103} (\bibinfo {year}
  {1961})}\BibitemShut {NoStop}%
\end{thebibliography}%

\end{document}